\documentclass[twocolumn,showpacs,preprintnumbers,amsmath,amssymb,pre,cite]{revtex4}
\usepackage{graphicx}
\usepackage{dcolumn}
\usepackage{bm}
\begin{document}

\preprint{/cond-mat.xxxxxxxx}

\title{Flow in linearly sheared two dimensional foams: from bubble to bulk scale}

\author{Gijs Katgert}
\author{Andrzej Latka}
\author{Matthias E. M\"{o}bius}%
\author{Martin van Hecke}
\affiliation{
 Kamerlingh Onnes Lab, Universiteit Leiden, Postbus 9504, 2300 RA
Leiden, The Netherlands}%

\date{\today}

\begin{abstract}
{We probe the flow of two dimensional foams, consisting of a
monolayer of bubbles sandwiched between a liquid bath and glass
plate, as a function of driving rate, packing fraction and degree of
disorder. First, we find that bidisperse, disordered foams exhibit
strongly rate dependent and inhomogeneous (shear banded) velocity
profiles, while monodisperse, ordered foams are also shear banded,
but essentially rate independent. Second, we introduce a simple
model based on balancing the averaged drag forces between the
bubbles and the top plate, $\bar{F}_{bw}$ and the averaged
bubble-bubble drag forces $\bar{F}_{bb}$, and assume that
$\bar{F}_{bw} \sim v^{2/3}$ and $\bar{F}_{bb} \sim (\partial_y
v)^{\beta}$, where $v$ and $(\partial_y v)$ denote average bubble
velocities and gradients. This model captures the observed rate
dependent flows for $\beta \approx 0.36$, and the rate independent
flows for $\beta \approx 0.67$. Third, we perform independent
rheological measurements of $\bar{F}_{bw}$ and $\bar{F}_{bb}$,
both for ordered and disordered systems, and find these to be
fully consistent with the forms assumed in the simple model.
Disorder thus modifies the exponent $\beta$. Fourth, we vary the
packing fraction $\phi$ of the foam over a substantial range, and
find that the flow profiles become increasingly shear banded when
the foam is made wetter. Surprisingly, our model describes flow
profiles and rate dependence over the whole range of packing
fractions with the same power law exponents --- only a
dimensionless number $k$ which measures the ratio of the
pre-factors of the viscous drag laws is seen to vary with packing
fraction. We find that $k \sim (\phi-\phi_c)^{-1}$, where $\phi_c
\approx 0.84$, corresponding to the 2d jamming density, and
suggest that this scaling follows from the geometry of the
deformed facets between bubbles in contact. Overall, our work
shows that the presence of disorder qualitatively changes the
effective bubble-bubble drag forces, and suggests a route to
rationalize aspects of the ubiquitous Herschel-Bulkley (power law)
rheology observed in a wide range of disordered materials.}
\end{abstract}

\maketitle
\section{Introduction}

Foams, which are dispersions of densely packed gas bubbles in a
liquid, exhibit an intricate mix of elastic, plastic and viscous
behavior reminiscent of the mechanics of other disordered
materials such as colloidal suspensions, granular media and
emulsions
\cite{cohen-addad,coussot,becu,kraynikannu,durian,denninpre74}.
When left unperturbed, foams jam into a meta-stable state where
surface tension provides the restoring force underlying their
elastic response for small strains
\cite{cohen-addad,kraynikannu,weaire}. Under continuous driving the
foam starts to flow, and the viscous dissipation that arises in
the thin fluid films that surround the gas bubbles becomes
important. Macroscopically, the steady state rheology of foams
exhibits shear thinning, and the stress $\tau$ as function of
strain rate $\dot{\gamma}$ is generally non-linear, often taking a
Herschel-Bulkley form: $\tau=\tau_Y+c_1\dot{\gamma}^\beta$, where
$\tau_Y$ denotes the yield stress, and where the viscous stress
$\tau_V\equiv \tau-\tau_Y$ scales nontrivially with the strain
rate $\dot{\gamma}$
\cite{cohen-addad,becu,larson,khan88,princen89,gopal,denninprl89,denkov1,katgert}.
In addition, in many situations, the flow is inhomogeneous and
localizes in a shear band
\cite{cohen-addad,denninprl89,katgert,debregeasprl87,denninprl93}.

In an earlier Letter \cite{katgert} we experimentally probed the
flow of disordered, bidisperse 2d foams which are trapped between
the fluid phase and a top-plate. The 2d nature allows for direct
imaging of the bubble dynamics and of the shear banded flow
profiles in this system. Combining measurements of the flow
profiles with rheological measurements, we established that
viscous interactions between neighboring bubbles scale differently
with velocity gradients than the effective viscous interactions at
the global scale. We captured the rate-dependent shear banding
exhibited by our system in a nonlinear drag force balance model.
Here we expand on these findings, discuss new results for the
effect of varying the wetness of the foam, and provide extensive
additional evidence to support our main conclusions.

To understand the rheology and shear band formation in our system,
three ingredients need to be described and combined appropriately:
{\em{(i)}} Interactions with the top plate. {\em{(ii)}} Local
bubble interactions. {\em{(iii)}} Disorder.

{\em Top plate ---} In recent years, a variety of studies have
addressed the formation of shear bands in (quasi) two-dimensional
foams, consisting of a single layer of macroscopic bubbles. Such
single layers can be made by freely floating the bubbles on the
surface of a surfactant solution (''bubble raft'')
\cite{denninprl89, bragg,denninpre73}, by trapping them between a top
glass plate and the surfactant solution (''liquid-glass'')
\cite{katgert,denninpre73,stanley, vaz, dolletpre}, or by trapping
them between two parallel glass plates (Hele-Shaw cell)
\cite{debregeasprl87}.

In a seminal paper by Debr\'{e}geas et al. \cite{debregeasprl87},
a bidisperse foam in Hele-Shaw Couette cell was sheared and narrow
shear banded flow profiles where obtained \cite{debregeasprl87}.
While initially it was believed that for slow flows, the effect of
the viscous drag forces exerted by the confining glass plates
would be negligible \cite{cohen-addad,debregeasjfm}, these drag
forces have turned out to be crucial. First, Couette experiments
in bubble rafts found completely smooth flow behavior
\cite{denninprl89}. Second, in experiments where a monodisperse
foam was linearly sheared with and without confining glass plate
on top \cite{denninpre73}, one observes smooth velocity profiles for
the bubble raft but highly shear banded flows for the liquid-glass
geometry. The precise connection between the drag forces due to
the confining plates and the occurrence of shear banding in
confined foams is still a subject of debate \cite{debregeasjfm,cheddadi}.

A simple continuum model that balances the top plate drags and the
inter-bubble drags (modeled with a Bingham constitutive relation)
captures both the rate independence and exponentially localized
shear bands seen in the linear liquid-glass cell
\cite{katgert,janiaud}
--- here we will build on and extend this model to capture the
experimentally observed nonlinear, rate dependent rheology of
disordered foams.

{\em Local interactions ---}  At the microscopic level, bubble
interactions are a combination of elastic repulsion, typically
harmonic for small deformations \cite{weitz,lacasse,brujic,dinsmore},
and nonlinear viscous drag forces \cite{katgert, denkov1, denkov2,
cantat, drenckhan, terriac, denkov3}. Such drag forces
arise when two bubbles slide past each other or when a bubble
slides past a solid boundary. The viscous drag forces originate in
the thin films that surround foam bubbles, and have recently
received renewed attention \cite{denkov1, denkov2, cantat,
drenckhan, terriac, denkov3}. Already for a single bubble sliding
past a solid wall, Bretherton showed that the drag force scales
nonlinearly with the bubble velocity \cite{
denkov1,terriac,bretherton}, and by analogy one would expect the drag forces
arising between sliding bubbles to be nonlinear also --- indeed
Denkov et al. recently suggested that a similar scaling applies to
the viscous drag force between bubbles \cite{denkov3}.

Here we measure these drag forces directly by rheological
experiments where two rows of ordered bubbles are sheared past
each other.

{\em Disorder ---}  Foam flows are disordered and intermittent at
the multi-bubble scale
\cite{durian,weaire,denkov1,katgert,janiaud,liuletter,mobius}.
For such disordered systems, the affine approach, where one simply
scales up local elastic or viscous interactions, often fails to
describe the macroscopic behavior --- this is by now well
established for shear deformations in granular and foam-like
systems \cite{makseprl1999,ohern,wouterEPL}, and a similar
picture is emerging from simulations of the flow of viscous
particles \cite{liuletter,olsson,hatano,langlois,remmers}. In the present work we
present strong experimental evidence for the failure of the affine
approach to describe drag forces in flowing systems.

\subsubsection*{Outline} In this paper, we describe an experiment in which
we have linearly sheared  a 2d foam and we
disentangle the roles of the top plate,
the local bubble interactions and the disorder, as well as the role of the wetness of the foam.

In section \ref{sec:exp} we describe our experimental setup. In
section \ref{sec:lin} we present experimental results for flow
profiles for a range of strain rates and span-wise widths of our
system. We find that the flow depends crucially on the applied
strain rate $\dot{\gamma}_a$: disordered, bidisperse foams exhibit
rate dependent flow profiles, which become increasingly
shear-banded for large $\dot{\gamma}_a$. We capture our findings
in a model in which the time-averaged drag forces between bubble
and top plate, $\overline{F}_{bw}$, and between neighboring
bubbles, $\overline{F}_{bb}$ are balanced. While the continuum
limit of our model is similar in spirit to \cite{janiaud},
the crucial new ingredient is nonlinear scaling laws for the wall
drag and the bulk stress ---  these nonlinear scalings are
essential for capturing the observed rate dependence.
\par
In section \ref{sec:rhe}, we probe the scaling of the drag forces
by independent rheological measurements, allowing us to directly
probe the role of disorder by comparing the rheology of small
ordered and larger disordered bubble rafts. We find
the averaged drag forces in the disordered foam to scale {\em
differently} from the local drag forces between individual
bubbles, which we have measured at high resolution and analyze in
a novel way.

In contrast, for monodisperse, ordered foams the local, averaged and top-plate drag forces all scale
similarly, causing rate-independent flows similar to those seen by
Wang et al. \cite{denninpre73}, and we discuss these in section
\ref{sec:ord}.

In section \ref{sec:pac} we further probe the connection between
the viscous drag forces at the bubble scale and the bulk viscous
forces by performing additional linear shear experiments over a range
of packing (air) fractions $\phi$. We find that the contribution
of averaged bubble-bubble drag forces vanishes algebraically as
$\phi-\phi_c$, when the packing fraction is decreased towards a
critical value $\phi_c$, which we identify with the (un)jamming
density --- $\phi_c \approx 0.84$.  We relate the vanishing of the
averaged bubble-bubble drag forces at $\phi_c$ to the vanishing
overlap between bubbles at unjamming.

The simple elastic interaction (typically harmonic for small
deformations \cite{weitz,lacasse,brujic,dinsmore}) and the
absence of solid friction make static packings of foam bubbles
eminently suited to compare to simulations of the popular soft
frictionless sphere model \cite{ohern,ellenbroek,makseprl2000}. Our
work illustrates the great potential of foams to elucidate the
{\em flow} behavior of simple systems near jamming
\cite{olsson,hatano,langlois,remmers}.

\section{Experimental details}\label{sec:exp}
In this section we describe in detail  a novel experimental setup to induce
linear shear flow in two-dimensional foams. We also
detail the analysis techniques used to extract velocity profiles, and
discuss measurements which show that coarsening and fluid drag can
be neglected.
\subsection{Setup}
We create foam bubbles on the surface of a reservoir of soapy
solution (of depth 3.5 cm), consisting of 80\% by volume demineralized water, 15\%
glycerol and 5\% Blue Dawn dishwashing agent (Proctor \& Gamble),
by bubbling nitrogen through the solution via syringe needles of
variable aperture. We measure the bath surface tension $\sigma$
with the pendant drop method \cite{hauser} and find $\sigma$ = 28
$\pm$~1 mN/m. We measure the dynamic viscosity $\eta$ with a
Cannon Ubbelohde viscometer and find $\eta$ = 1.8$\pm$~0.1 mPa.s.
\par
Fig.~1 shows our experimental setup: the bubbles are contained
inside an aluminum frame (400x230 mm) which is leveled with the
liquid surface and which supports glass top plates to which the
bubbles bridge once they are in place. The top plates consist of
three adjacent glass plates with slits to accommodate two PMMA
wheels of radius 195 mm and thickness 9.5 mm which drive the flow.
The vertical gap between the liquid surface and the glass plates
can be varied to control the packing fraction of bubbles $\phi$.
\par
The wheels, which are grooved to provide a no-slip boundary for
the bubbles, can be lowered into and raised out of the bath
through the slits. The wheels are connected to two Lin Engineering
stepper motors, each driven by micro-stepping driver, and are
rotated in opposite directions. At any point along the line where
the wheels contact the foam bubbles the horizontal component of
the driving velocity is  a constant (see Fig.~1b).
\par
\begin{figure}[t]
\begin{center}
\includegraphics[width=\columnwidth]{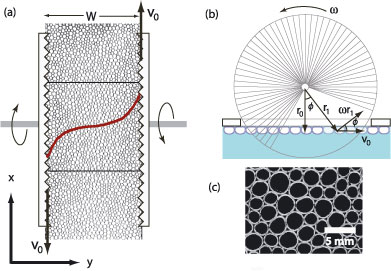}
\caption{(Color online) (a)~Schematic topview of the experimental
setup. $W$ represents the gap width and the two horizontal lines
indicate the edge of the region over which the velocity profiles
are calculated. The red curve depicts one such profile.
(b)~Sideview of the shearing wheels. The slits in the glass plate
are drawn for clarity. That the in-plane component of the motion
of the boundary is constant can be seen as follows: by trivial
geometry, we obtain that  $v_0 \!=\!\omega r_1 \cos \phi$, but
since $r_1 =\frac{r_0}{\cos\phi}$, at any point along the along
the contact line of 230 mm, the layer of bubbles is sheared with a
driving velocity $v_0 =\omega \frac{r_0}{\cos\phi} \cos \phi
=\omega r_0$. (c)~Experimental image of part of the foam, the
scalebar represents 5 mm.} \label{setup}
\end{center}
\end{figure}

We obtain our data  from the central 60 mm of the shearing region
--- marked by the horizontal lines in Fig.~\ref{setup}(a)--- to
avoid effects caused by the recirculation of the foam at the edges
of the wheels. In this central part no motion is observed due to
the vertical component of the radial velocity. At the edges of the
slits, bubbles do leave the system, while being pinned to the
wheels. This does not result in holes in the foam layer, either
because at high driving velocities the bubbles reenter the system
before rupturing while traveling on the wheel, or because at low
velocities  bubbles from outside the shearing region are pushed
inwards due to the bubble surplus at the edges. The resulting
driving velocity gives rise to a global strain rate $\dot{\gamma}
=2v_0/W$, where $W$ denotes the gap between the wheels, which we
vary between 5 and 10 cm.
\par
\subsection{Imaging and Analysis}
\begin{figure}[th]
\includegraphics[width=80mm]{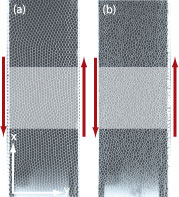}
\caption{(Color online) Images of sheared regions for both (a)
monodisperse and (b) bidisperse foams.  Shear is indicated by the arrows. The highlighted area is where data analysis is performed on. } \label{image}
\end{figure}

We wish to characterize the average flow in the $x$-direction as a
function of the span-wise coordinate $y$. The average velocity
profiles are obtained from a series of images which we record with
an 8 bit Foculus BW 432 CCD camera (1280x1024 pixels) equipped
with a Tamron 28-300 telezoom objective. In the images, 1 pixel
corresponds to approximately 0.1 mm. To optimize the brightness
and obtain images in which the bubbles appear as circles,
the foam is lit laterally by two fluorescent tubes, each driven by
high frequency ballasts to prevent flickering in the images. The
bottom of the reservoir is covered with a black plate to improve
contrast. Typical images are shown in Fig.~\ref{image}.
\par
The frame rate is fixed such that the displacement at the wheels
is fixed at 0.15 mm between frames. Since the flow is strongly
intermittent, with large fluctuations in the bubble displacements,
we take 1000 frames per run, corresponding to a strain of 4 for a
5 cm gap, as we are interested in averaged velocity profiles. We
pre-shear the system before taking data, so that a steady state is
reached.
\par
We obtain the velocity profiles both through particle tracking and
a Particle Image Velocimetry-like technique, where for each
$y$-value, we calculate the cross-correlation $(C_n)^2$ between
the corresponding image line in the $P_n(x)$ of length $m$ and the
same image line  $P_{n+1}(x)$ in the next frame shifted by an
amount $\tau$:
\begin{equation}
 (C_n(\tau))^2 = \sum_{i=0}^{m-\tau}  P_n(i)P_{n+1}(i+\tau).
\end{equation}

We can then proceed in two ways. The first method is to add
up all cross-correlations from all frames for each $y$-value, and calculate the
average displacement $\Delta x$ per frame by fitting a parabola $p_n(\tau)$ to
the resulting sum of cross-correlations and taking the peak value of that parabola:
\begin{equation}
\Delta x(y) = {\bf \rm max}\left(\sum_{n=0}^{999} p_n(\tau) \right).
\end{equation}
In the second method we fit a parabola to each cross-correlation separately
and obtain the average displacement by averaging the maxima of all individual parabolas:
\begin{equation}
\Delta x(y) =\langle{\bf \rm max}\left(p_n(\tau)\right)\rangle .
\end{equation}
\par
By comparing to average velocity profiles obtained by particle tracking \cite{mobius},
we find that the latter procedure gives the closest match to the tracking velocity profiles,
and we have employed this procedure throughout. We thus obtain both spatially (in the $x$-direction)
and temporally averaged velocity profiles.  Despite the intermittent character of the flow, we obtain smooth
reproducible velocity profiles.
\subsection{Coarsening and fluid drag}
To characterize the amount of coarsening we measure the bubble
size distribution by measuring the surface area of the bubbles in
the images. We obtain well defined size distributions which show
little coarsening over the duration of the runs, which corresponds
to about 2 hours (Fig.~3a).
\begin{figure}[th]
\begin{center}
\includegraphics[width=80mm]{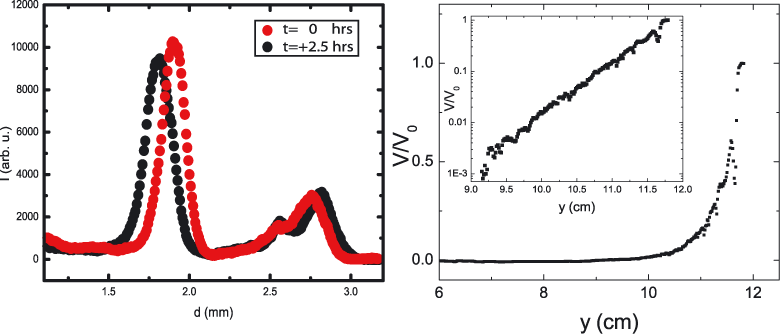}
\caption{(a) Size distribution and coarsening over the duration of an experimental
run for bidisperse foams. (b) Flow at the
liquid surface in the absence of bubbles, as imaged by depositing
silver powder. Inset: same profile on lin-log scale, showing
exponential decay away from the boundaries.} \label{bulkflow}
\end{center}
\end{figure}
\par
We have checked that the drag on the foam bubbles due to flow  of the bulk liquid underneath
is negligible by measuring the velocity profile of
bubbles floating on a very shallow layer of bulk fluid. In this case the fluid surface velocity
is decreased due to the no-slip boundary condition at the reservoirs' bottom. We do not, within experimental uncertainty,
observe a change in the experimental velocity profiles in this geometry.
\par
We furthermore measure the velocity profile of the liquid surface
itself at the same fluid level as in the foam experiments
by imaging the flow of silver particles that
were sprinkled on the liquid surface, see Fig.~\ref{bulkflow}b. We
observe a steeply decreasing velocity profile at the fluid
surface, which implies that even if the fluid drag were of the
order of the other drags acting on the bubbles, it would not
significantly alter the flow profiles except near the wheels.
\par
We thus conclude that the bubble size distribution is essentially
constant during an experimental time frame, and that the dominant
drag forces are those between bubbles and top plate, and those
between contacting bubbles.
\section{Linear shear of two dimensional foams}\label{sec:lin}
\begin{figure*}[t]
\begin{center}
\includegraphics[width=17cm]{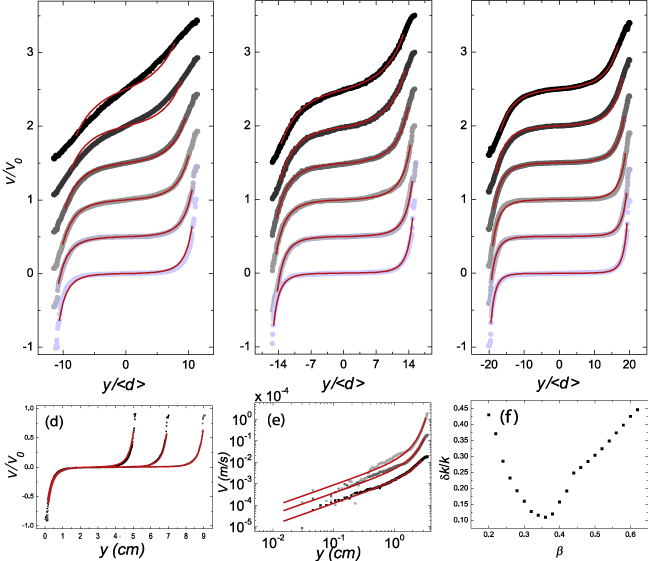}
\caption{(Color online) (a-c) Flow profiles for a gap width $W$ =
5 (a) 7 (b) and 9(c) cm. From black to light grey, $v_0=$ 0.026
mm/s, 0.083 mm/s, 0.26 mm/s, 0.83 mm/s, 2.6 mm/s and 8.3 mm/s. For
all gap widths we observe that the localization near the driving
wheels  increases for increasing driving velocity. For clarity the
profiles are each offset vertically by 0.5$\times v/v_0$. Solid
red curves: fits to the drag force balance model of section III B.
(d) Profiles at 2.6 mm/s for all three gap widths. Regardless of
the gap width all profiles decay at the same rate. (e) Examples of
profiles and fits on a log-log plot, highlighting the linear tails
of the profiles. $v_0=0.026,~0.26,~2.6$~mm/s, $W=5$~cm. (f) As
explained in section III C \emph{1}, the minimum in $\beta$ is
found by calculating the variance $\sqrt{\langle \delta k ^2
\rangle/k^2}$ of $k$ over all 18 profiles depicted in (a-c). The
minimum in the variance is seen at $\beta=0.36$ --- see section
III C.} \label{ratedep}
\end{center}
\end{figure*}

In this section we explore the rate dependent shear flows in our
system experimentally. By fitting our experimental data to a
nonlinear drag force balance model, we deduce the dependence of
the averaged bubble-bubble and bubble-wall drag forces as function
of the local strain rate and velocity.

\subsection{Flow of disordered foams}
We measure averaged velocity profiles in disordered
two\-dimen\-si\-o\-nal foams. These foams are produced by bubbling
a fixed flow rate of nitrogen through syringe needles of 2
different inner diameters, such that bubbles of 1.8 $\pm 0.1$  and
2.7 $\pm 0.2$ mm result (at 59-41 number ratio). The bubbles are
gently mixed with a spoon until a disordered monolayer results.
For gap widths of 5, 7 and 9 cm, we drive the foam at 6 different
velocities, spanning 2.5 decades: $v_0 =$ 0.026, 0.083, 0.26,
0.83, 2.6 and 8.3 mm/s.

Note that we perform the sweep in driving velocities from fast to
slow and that we pre-shear the system for one full wheel rotation,
to start with bubbles covering the wheel \cite{foot1} and ensure
that we have reached a steady state. To fix the packing fraction,
we fix the gap between  glass plate and liquid surface at 2.25
$\pm$~0.01 mm. We have measured (see section VI) that for this gap
the packing fraction is $\phi =0.965 \pm 0.005$.
\par

Results are plotted in Fig.~\ref{ratedep}(a-c): the profiles exhibit shear banding,
and for all gap widths the profiles become increasingly shear banded at increasing driving velocities.
The slowest runs at $W$ = 5 cm yield essentially linear velocity profiles.
We suggest that these shapes are due to the small gap width, which results in overlapping shear banded
profiles resembling a linear profile, and in what follows, we will present a model that supports
this conclusion.
\par
In Fig.~\ref{ratedep}(d) we plot velocity profiles for a driving velocity of 0.26 mm/s for all
three gap widths together, which clearly show that for all widths, the velocity profiles decay similarly.
Fig.~\ref{ratedep}(d) thus suggests that in this experiment the driving velocity at
the edges, instead of the overall shear, sets the velocity profiles, and that the
local response to forcing will provide the key towards understanding the shape
of these profiles. Note finally that the profiles do not exhibit significant slip with respect
to shearing wheels, except for the fastest runs, where the slip is less than 20 \%.

\subsection{Model}
We now propose a model to account for the shear banding behavior discussed above,
by considering the balance of the averaged viscous drag forces.
\subsubsection{Drag forces on individual bubbles}
The drag force on a single bubble that slides past a solid wall was first investigated by Bretherton \cite{bretherton} and has
recently received renewed attention \cite{denkov1,denkov2,cantat,drenckhan,terriac,quere}.
The crucial finding is that
$F_{bw}$, the drag force per bubble sliding past a solid wall, scales as
\begin{equation}
F_{bw} = f_{bw}(Ca)^{2/3} = f_{bw}\left(\eta v/\sigma \right)^{2/3},
\label{fbb}
\end{equation}
with $\eta$ the bulk viscosity, $\sigma$ the surface tension, $f_{bw}$ a constant with dimensions
of force and $Ca$ the capillary number. Typically $f_{bw} \propto \sigma r_c$, with $r_c$ the radius of the deformed
contact between bubble and wall \cite{quere}.
For bubbles in a soapy solution, the 2/3 scaling with $Ca$ only holds for surfactants that
are mobile \cite{denkov1}. Results from \cite{Koehler} strongly indicate that this is indeed the case
for our surfactant Dawn, as we will confirm below.
\par
The drag force between 2 bubbles sliding past each other,
$F_{bb}$, has not received much attention up to now, although
\cite{kraynik} provides indirect evidence that it scales like
$F_{bb} \propto \left(\eta \Delta Ca \right)^{\zeta}$, with $
\Delta Ca \equiv \eta \Delta v/\sigma $. In a very recent Letter
it is explicitly shown that, for ordered bubble motion $F_{bb}$
scales indeed as $(\Delta Ca)^{\zeta}$ \cite{denkov3}. The authors
find $\zeta =0.5$, although various physico-chemical
peculiarities, as well as the range of $Ca$ one measures in, can
alter this exponent.
\par
Taking all of this into consideration, it seems reasonable to assume that:
\begin{equation}
F_{bb}= f_{bb}\left(\eta \Delta v/\sigma \right)^{\zeta}.
\label{fbw}
\end{equation}
While the dissipation leading to $F_{bw}$ occurs at the perimeter
of the flattened facet \cite{denkov1} --- hence the prefactor
$f_{bw} \propto \sigma r_c$ ---  $f_{bb}$ scales $ \propto \sigma
\kappa_c^2$, with $\kappa_c$ the radius of the deformed contact
between bubbles, thus reflecting the different physical mechanism
behind this scaling \cite{denkov3}.
\begin{figure}[h]
\begin{center}
\includegraphics[width=80mm]{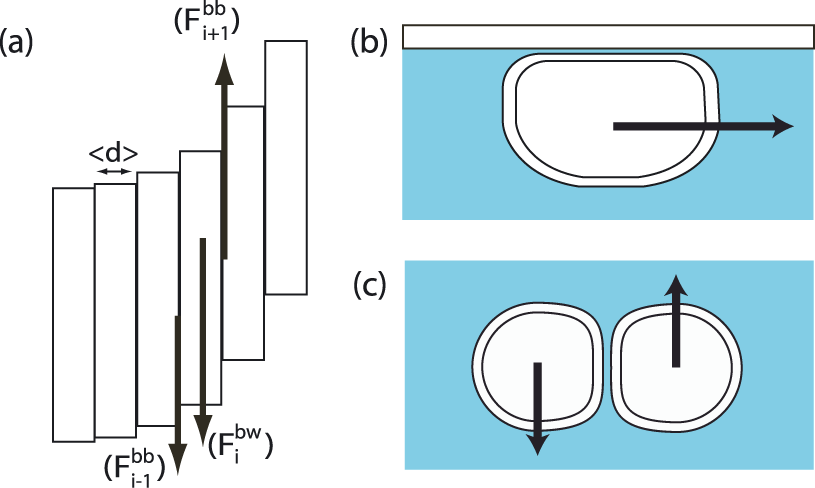}
\caption{(Color online) Illustration of drag balance model. The shear region is divided in lanes labeled
$i$ which all experience drag forces due to the top plate and due to both neighboring lanes. (b-c)
Illustration of the films around which the viscous drag forces act.} \label{model}
\end{center}
\end{figure}
\subsubsection{Stress balance}
We divide our shearing region in lanes labeled $i$ and assume that on every lane  the \emph{time-averaged} top plate drag per bubble
$\overline{F}_{bw}^i$ balances with the time-averaged viscous drag per
bubble due to the lane to the left ($\overline{F}_{bb}^{i}$) and
right ($\overline{F}_{bb}^{i+1}$), see Fig.~\ref{model}:
\begin{equation}
\overline{F} \vspace{0.1 cm}_{bb}^{i+1} - \overline{F} \vspace{0.1 cm}_{bw}^i -\overline{F} \vspace{0.1 cm}_{bb}^{i}=0.
\label{forcebal}
\end{equation}
We assume that the averaged drag forces scale similarly to the local drag forces.
For the averaged bubble-wall forces we assume:
\begin{equation}
\overline{F}\vspace{0.1 cm}_{bw}^i = f_{bw}(\eta v^i /\sigma)^{2/3}, \label{HBI}
\end{equation}
while for the averaged bubble-bubble drag forces we assume:
\begin{eqnarray}
\overline{F}\vspace{0.1 cm}_{bb}^i &=& f_Y+f_{bb}\left[(\eta/\sigma)(v^i-v^{i-1})\right]^{\beta}~,\label{HBII}\\
\overline{F}\vspace{0.1 cm}_{bb}^{i+1} &=& f_Y+f_{bb}\left[(\eta/\sigma)(v^{i+1}-v^{i})\right]^{\beta}
~.\label{HBIII}
\end{eqnarray}
Here $f_{bw}$ and $f_{bb}$ are material parameters with dimension
of force, which will be measured by rheometry in section IV below.
Finally, $f_Y$ represents a yield force in the inter-bubble drag,
to remain consistent with rheometrical data presented later on and
to reflect the elastic barrier bubbles have to overcome before
they slide past each other. Note that the velocities $v^i$ denote
the \emph{averaged} velocities in the $x$-direction --- the
crucial assumption is that the relation between the averaged drag
forces and the averaged velocities is simple and can be expressed
by a single power law.

We do not know  if and how the conjectured forms for the averaged
forces can be derived from the non-averaged forces
Eqs.~(\ref{fbb}) and (\ref{fbw}) since due to the intermittent and
disordered bubble motion, the instantaneous bubble velocities are
fluctuating and not necessarily pointing in the $x$-direction. For
example, there is no a priori reason for the exponents $\zeta$ and
$\beta$ to be equal and in fact our data strongly indicate that
they are not. The best justification for
Eqs.~(\ref{HBI}-\ref{HBIII}) is a posteriori --- the resulting
model describes the data well. Note that the bars
in Eqs.~(\ref{HBI}-\ref{HBIII})  express an average over disorder,
in the sense that these quantities are measured in highly
disordered, intermittent flows.
\par
Inserting the expressions from Eqs.~(\ref{HBI}-\ref{HBIII}) into
Eq.~(\ref{forcebal}) and defining $k=f_{bw}/f_{bb}$ we arrive at:
\begin{equation}
k \left( \frac{\eta v^{i}}{\sigma}\right)^{2/3} =
\left(\frac{\eta}{\sigma}\right)^{\beta}
\left[(v^{i+1}-v^{i})^{\beta} - (v^i-v^{i-1})^{\beta}\right]~.
\label{modeleq}
\end{equation}
Note that the yield drag contributions $f_Y$ cancel, which is a
particular advantage of the linear geometry we work in. The model
predicts flow profiles for arbitrary width and driving rate, once
the parameters $\beta$ and $k$ are fixed.
\subsection{Fits}
\subsubsection{Procedure}
We compare all 18 runs to solutions of the model. We focus on the
central part of the data where $|v|<3/4v_0$ to avoid the edge
effects near the shearing wheels (for instance the bumps in the
low-velocity profiles in Fig.~\ref{ratedep}(a) and the slip with
respect to the wheel in the fast runs). We numerically integrate
Eq.~(\ref{modeleq}) from $y=0$, where $v=0$, to the $y$ value for
which $v=3/4 \cdot v_0$, while keeping $\beta$ and $k$ fixed. The
drag force balance should govern the shape of the velocity
profiles for all driving rates and gap widths. Therefore we
determine for fixed $\beta$ the $k$ values that fit the flow
profiles best. The $k$ values exhibit a systematic variation that
depends on the value of $\beta$. We quantify this variation by
computing the relative variance $\sqrt{\langle \delta k^2 \rangle/
k^2}$ and by repeating the procedure for a range of $\beta$, we
obtain a plot of the variance as a function of $\beta$, see
Fig.~\ref{ratedep}(f). From this graph, we determine the value for
which the variance is minimized as $\beta=0.36 \pm 0.05$.

\subsubsection{Results}
Fixing now $k=3.75$ and $\beta=0.36$, we  capture the shape of all
data sets with high accuracy . The resulting model profiles are
plotted in Fig.~\ref{ratedep}(a-c), and we see that for these
values all velocity profiles are adequately fitted except for the
slowest runs at $W=5$~cm. We attribute these deviations for small
$W$ to the observation that edge effects extend further into the
shearing region for small gaps.
\par
Note that the model profiles exhibit linear tails, see
Fig.~\ref{ratedep}(e), and that the experimental velocity profiles
exhibit approximately the same behavior. We conclude that both the
experimental and model profiles do not decay exponentially, in
contrast with results found in previous studies \cite{debregeasprl87,denninpre73}.
\par
\begin{figure*}[t]
\includegraphics[width=18cm]{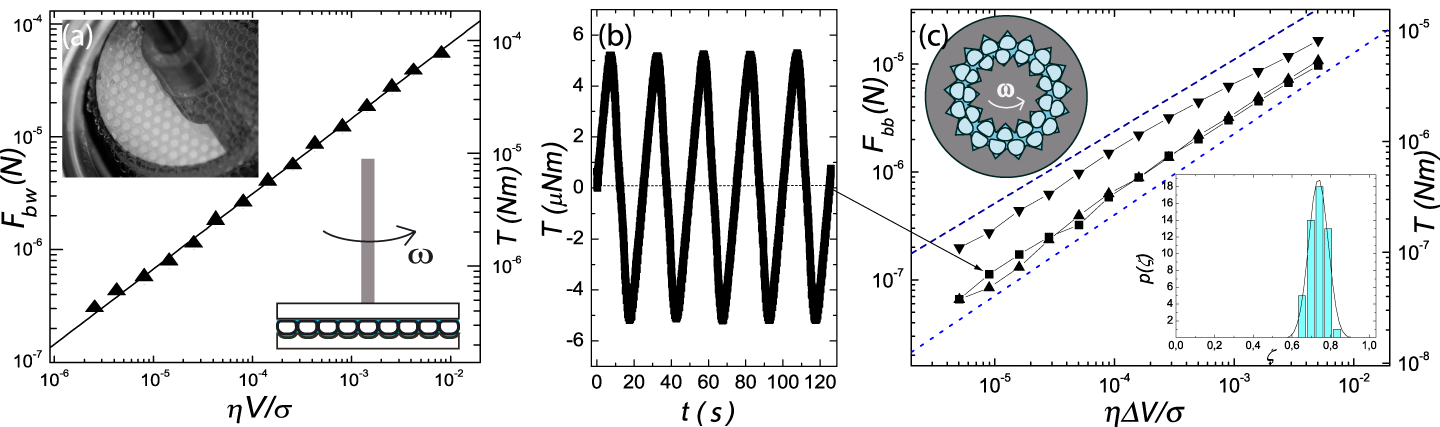}
\caption{(Color online) (a) Drag force per bubble exerted on
smooth rotated plate as a function of $Ca$, probed by the total
drag force of a pinned layer of bubbles on a rotating top plate.
The solid line represents $F_{bw} = 0.0015 \pm 0.0001 \cdot \left (\eta
v/\sigma \right)^{2/3}$. The upper inset shows a close-up
photograph of the rheometrical tool used to measure the
bubble-wall drag: the reflection of the flattened facets of radius
$r_c$ used to extract $R_0$ can be seen clearly. The lower inset
shows a side view  of the experimental geometry. (b) Raw torque for
ordered and commensurate lanes of bubbles sliding past each other.
Notice the huge fluctuations with respect to the mean indicated by
the horizontal line . The
average of the raw data corresponds to the data point in (c)
indicated by the arrow. (c) Torque averaged over an integer
number of rearrangements as a function of $\Delta Ca$ for the
commensurate case (40 bubbles on inner wheel, 40 bubbles on outer
wheel)($\blacksquare$), incommensurate case 41/40
($\blacktriangledown$), incommensurate case 44/40
($\blacktriangle$). Dashed lines indicate $\zeta=0.67$ resp.
$\zeta=0.75$. Upper inset shows a schematic picture of the
rheometrical geometry, lower inset shows a histogram of the
extracted values of the exponent $\zeta$. The width of the bin
indicates the error in $\zeta$.} \label{andydragforces}
\end{figure*}
\subsection{Continuum Limit}
The continuum limit of Eq.~(\ref{modeleq}) can be written as:
\begin{eqnarray}
f_{bw}\left( \frac{\eta v}{\sigma} \right)^{2/3} \langle d \rangle ^{-1} &=& \frac{\partial \tau}{\partial y}~, \\
\tau = \tau_Y + f_{bb} \left( \frac{\eta \left<d\right>\dot{\gamma}}{\sigma}\right)^{\beta},&~&~~~ \beta=0.36~. \label{herschey}
\end{eqnarray}
Hence, the top plate drag can be considered as a body force and the
inter-bubble drag force as the divergence of a shear stress $\tau$,
where $\tau_Y$ is an undetermined yield stress. Eq.~\ref{herschey} is the constitutive equation for a Herschel-Bulkley fluid \cite{hb}, and we can now associate the averaged bubble drag force scaling at the local level with the power law
scaling of the viscous stress in the Herschel-Bulkley model.
\par
Note that $\beta = 0.36$ is similar
to the power law index $n=0.40$  found for the bulk rheology of three-dimensional mobile
foams \cite{becu, denkov1} and to the values $n=0.33$ and $n=0.45$ found
for two dimensional bubble rafts in a Taylor-Couette geometry in \cite{denninprl93}.
\par
The fact that the yield stress does not play a role for our
velocity profiles  can now be understood in two ways: at the
continuum level, since it is  a constant it vanishes after taking
the divergence of the shear stress, at the bubble level, even
though we include a yield force in Eqs.~(\ref{HBI})-\ref{HBII}),
the contributions from both neighboring lanes cancel in
Eq.~(\ref{modeleq}). Finally, notice that the continuum equations
can easily be solved in terms of hypergeometric functions
\cite{btighe}.

\section[Rheometry of viscous forces in 2D foams]
{Rheometrical determination of viscous forces in two-dimensional
foams}\label{sec:rhe}

In this section we will investigate the viscous forces that act at
the bubble scale by rheometry, to test and validate the
assumptions for the scaling of the bubble-wall drag and the
viscous friction inside the foam expressed in
Eqs.~(\ref{fbb}-\ref{fbw}).  We use an Anton Paar DSR 301 stress
controlled rheometer, which can also operate in strain controlled
mode. We use the rheometer in strain controlled mode to
investigate $F \vspace{0.1cm} _{bw}$. Moreover, we compare
measurements, which reflect the actual drag force at the single
bubble level ($F\vspace{0.1cm} _{bb}$),  with measurements of the
averaged viscous drag force on a bubble in a disordered flow of
foam ($\overline{F}\vspace{0.1cm} _{bb}$).

\subsection{Bubble-wall drag}
We directly measure  the bubble-wall friction for foam bubbles
produced from the soap solution presented above, with a method
that was introduced in \cite{denkov1}. We load a monolayer of
bubbles between two PMMA plates of radius
$R_P$ = 2 cm. The bubbles are pinned to the lower plate by means
of a hexagonal pattern of indentations of size ${\mathcal O} (d)$,
and can slip with respect to the smooth upper plate which is
connected to the rheometer head, see lower inset of
Fig.~\ref{andydragforces}(a). We measure the torque $T$ exerted by
the bubbles as a function of the angular velocity $\omega$ of the
smooth plate.
\par
We convert $T(\omega)$ to $F \vspace{0.1cm}_{bw}(Ca)$ in the following way:
each bubble exerts a wall stress
$\tau_{w} =F \vspace{0.1cm} _{bw}/ \pi R_0^2$ on the smooth plate.
We integrate the contribution to the torque of this wall stress over the plate:
\begin{equation}
T = \int^{R_P}_0 \tau_{w} r 2 \pi r dr = \int^{R_P}_0 \frac{F \vspace{0.1cm} _{bw}}{R^2_0} 2 r^2 dr. \label{integ}
\end{equation}
If we now assume that $F \vspace{0.1cm} _{bw} \propto  \left[ Ca \right]^{\alpha} = \left [ \frac{\eta \omega r}{\sigma} \right ]^{\alpha}$,
we can immediately read of from the data that $\alpha = 0.67$, see Fig.~\ref{andydragforces}(a), so inserting this expression in the integral Eq.~(\ref{integ}) yields:
\begin{equation}
T= \frac{2 F \vspace{0.1cm} _{bw} R_p^{3.67}}{3.67 R_0^2}.
\end{equation}
Since the bubbles are flattened during the measurement, we can
only measure $R_0$ through the flattened facet $r_c$ by looking at the reflection of the deformed
facet, see the upper inset of Fig.~\ref{andydragforces}(a). We
find $r_c = 1.59 \pm 0.05$ mm. As the bubble radius is
smaller than $\kappa^{-1}$ we can express $R_0$ in terms of $r_c$
through $R_0^2 = \sqrt{\frac{3}{2}} r_c \kappa^{-1}$ \cite{quere}.
Note that this derivation of $r_c$ in terms of $R_0$ hinges on the
assumption that the bubbles are not too deformed, which is not
obvious in the rheometrical geometry, but for lack of a more
precise relation we use it. We finally rescale the horizontal axis
by  multiplying $\omega$ with $\eta R_p /\sigma$. The resulting
curve is plotted in Fig.~\ref{andydragforces}(a): over our
measurement range (more than three decades) $F_{bw} \propto
[Ca]^{2/3}$.

\subsection{Bubble-bubble drag}
\subsubsection{Drag at the bubble scale}
To measure the power law scaling of the inter-bubble drag we
measure the torque exerted by a foam driven at a strain rate
$\dot{\gamma}$ in a cylindrical Couette geometry, which consists
of an inner driving wheel, connected to the rheometer head,
rotating inside an outer ring. The rheometrical experiments are
performed with bubble rafts, i.e. foams that are not confined by a
top plate, as the additional stresses due to the wall would
disturb a clean rheological measurement.
\begin{figure*}[ht]
\includegraphics[width=18cm]{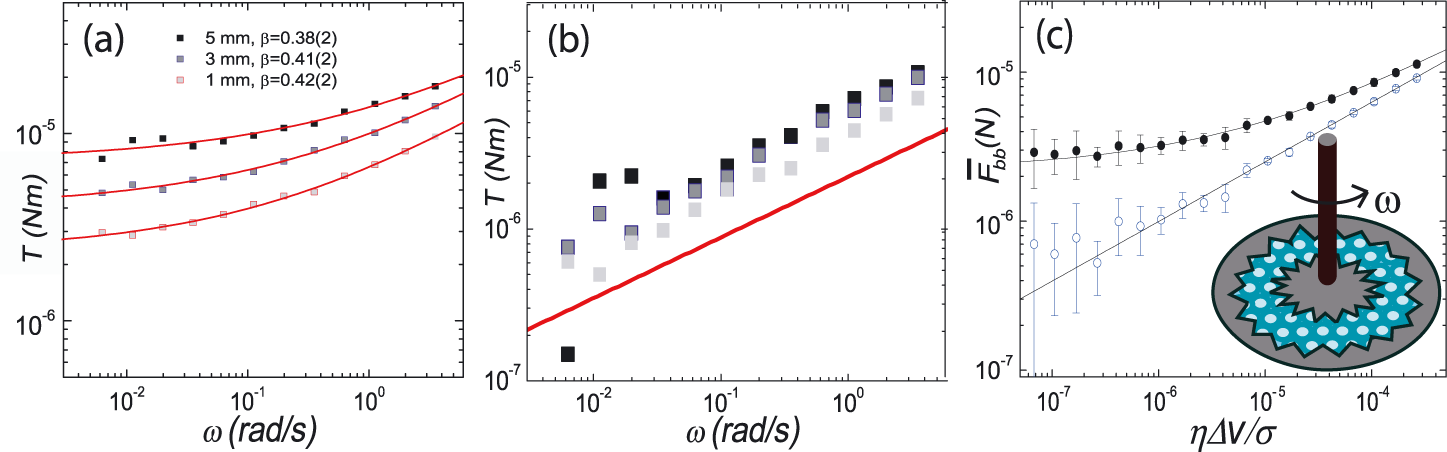}
\caption{(Color online) (a) Torque exerted  on the inner wheel by
a \emph{monodisperse} foam in a Taylor-Couette geometry with $r_i=25$mm, $r_o=70$mm,
for different bubble sizes as
indicated. Fits are to a Herschel-Bulkley model, and power law
indices $\beta$ from these fits are shown in the graph as well.
Surprisingly, the yield stress increases with increasing bubble
size --- see text. (b) Same data as in (a) with the yield torque
from the fit subtracted. The solid line is a power law with index
0.4. (c) Averaged drag force per bubble in a \emph{bidisperse},
disordered foam. The foam is sheared in a Couette cell of inner
radius 5 cm, outer radius 7 cm (hence a gap of 9 bubble diameters)
without a top plate, see inset. We obtain
$\overline{F}_{bb}=f_Y+f_{bb} (\Delta Ca)^\beta$, with the yield
threshold $f_Y \approx 2.2 \pm 0.5 \times 10^{-6}$~N, $f_{bb}
\approx 2.5 \pm 0.9 \times 10^{-4}$~N and $\beta=0.40 \pm 0.02$
(solid line). Open circles are the same data with the yield torque
obtained from the fit subtracted, which are well fit by a pure
power-law with exponent 0.4 (dashed line).} \label{slip}
\end{figure*}
\par
Both boundaries are grooved to ensure a no slip boundary for the
bubbles, of which a monolayer floats in the shearing region. We
start with measuring $\overline{F} \vspace{0.1 cm} _{bb}$ for the
ordered case by keeping the gap between the cylinders such that
exactly two layers of bubbles fit in, see the upper inset of
Fig.~\ref{andydragforces}(c). The inner radius ($r_i$) is 2.5 cm
and the outer radius ($r_o$) is 3.0 cm. We deposit bubbles of
2.2~mm diameter in the grooves, make sure that all bubbles are
strictly pinned and remain in their groove, and vary the rotation
rate $\omega$ of the inner cylinder over 3 decades while measuring
the torque averaged over an integer number of rearrangement
events, see Fig.~\ref{andydragforces}(c).

We multiply $\omega$ by $\eta r_{i}/\sigma$ to rescale the
dimensionless velocity difference and we divide the torque by
$r_{i}$ and the number of bubbles pinned at the inner wheel (e.g.
40) to obtain the averaged bubble-bubble drag force per bubble in
the ordered case.

We use three different inner wheels: one with 40 grooves, a second
with 41 grooves and a third with 44 grooves. Since the number of
grooves in the outer ring is fixed at 40, this allows us to
investigate the differences between commensurate and
incommensurate numbers of bubbles in the grooves.
\par
For the commensurate case, the result is plotted in
Fig.~\ref{andydragforces}(b): All bubbles rearrange simultaneously
and thus the signal reflects the torque exerted on a single
bubble, amplified by a factor of 40. The elastic barrier that
bubbles have to cross before rearranging is clearly visible in the
signal. As a result, the torque oscillates tremendously.
Nevertheless, the force per bubble averaged over many such events
scales with the dimensionless velocity difference as a power law
with index 0.7, see Fig.~\ref{andydragforces} (c). This value is
remarkably close to the exponent found for the bubble-wall drag.
 For these ordered lanes, no signs of a yield plateau are observed in the time averaged signal, and we believe this is due to the fact
that all elastic energy that is stored in the bubble deformation is released
after yielding, so that one measures purely the viscous drag.

\par
For the incommensurate runs, the raw signal looks more complex, as
rearrangements do not occur simultaneously for all 41 or 44
bubbles. The resulting power-law exponents for the averaged drag
forces are, however, close to the one observed for the
commensurate case. In fact, if we repeat the measurements for both
commensurate and incommensurate bubble numbers a multitude of
times and fit $Ca^{\zeta}$ to the averaged $F_{bb}$, we find a
distribution of $\zeta$-values around $\zeta =0.73$, see lower
inset of Fig.~\ref{andydragforces}(c). The binsize is similar to
the errorbar on each individual measurement. \subsubsection{From local to bulk viscous drag}

We observe that the scaling exponent for the viscous drag at the
bubble scale, $\zeta$, differs markedly from the scaling exponent
$\beta$ of the drag forces inside the bulk foam as extracted from
the velocity profiles, e.g., $\zeta \approx 0.70$ vs.
$\beta=0.36$. We hypothesize this is due to the disordered flow in
the foam and will provide rheological evidence in what follows.
\par

To perform rheological measurements of the drag forces, we employ
a Couette cell which has an outer ring of radius $r_0 =7$~cm, such
that more layers of bubbles can fit inside the cell. We first will
perform measurements on disordered packings of monodisperse
bubbles of three different sizes (1, 3 or 5 mm). We observe that
the foam deviates substantially from hexagonal packing during flow
because the inner radius $r_i=2.5$~cm is small, and the curvature
is large. We thus induce disorder through geometry.

The resulting measurements show clear yield stress behavior and
can be excellently fit by the Herschel-Bulkley model, yielding for
all bubble sizes $\beta \approx 0.4$, which is markedly lower than
the 0.70 found for the drag force in ordered lanes above, and close
to the 0.36 extracted from the velocity profiles (see
Fig.~\ref{slip}a-b). The observed  stress plateau at low strain
rates increases with increasing bubble radius, contrary to the
intuition that the yield stress is set by the Laplace pressure and
should hence scale inversely to the bubble radius. We tentatively
attribute this to the deformation of the bubbles through capillary
effects, which are larger for larger bubbles and hence lead to a
relatively larger contact size between the bubbles.

In order to further establish a connection between the
rheometrical data and the model, we now turn to a geometry with a
large inner wheel to increase the measured signal ($r_i =5$ cm and
$r_o = 7$ cm), and measure the torque exerted on the inner wheel
by a \textit{bidisperse} foam with the same bubble sizes as in the
linear shear experiment. We obtain a clear confirmation that
indeed the disorder changes the power law scaling of $\overline{F}
\vspace{0.1cm} _{bb}$: we again reproducibly measure
Herschel-Bulkley behavior with power law index $\beta \approx
0.40$, as can be seen in Figs.~\ref{slip}(c).

To convert torques to $\overline{F} \vspace{0.1cm} ^{bb}$, we
divide the torque by the number of bubbles and $r_i$. Since our
outer rough boundary forces the bubble velocity to zero, we can
rescale the angular frequency to the dimensionless velocity
difference $\eta \Delta v/ \sigma$ by assuming a linear velocity
profile across the gap, decaying from $\omega r_i$ to 0. The gap
width is approximately $9\langle d \rangle$ and hence we can
estimate $\Delta v$. We extract from the rheological measurements
an estimate for the ratio $k= f_{bw}/f_{bb} \approx 5.5 \pm 0.5$.
This is remarkably close to the value $k=3.75 \pm 0.5$ extracted from the
velocity profiles, given the crude estimates used in converting torques to bubble-bubble drag forces
in the rheometrical data --- we have oversimplified the shape of the velocity profile in the disordered Couette rheometry, which is neither linear, nor rate independent.
\subsection{Interpretation}
The drag forces exerted on the bubbles by the top plate, which at
first sight might be seen as obscuring the bulk rheology of the
foam, enable us to back out the effective inter-bubble drag forces
and constitutive relation of foams from the average velocity
profiles. To further appreciate this fact, note that our model
yields linear velocity profiles regardless of the exponent $\beta$
if the body force due to the wall drag is zero. This is consistent
with earlier measurements by  Wang et al. \cite{denninpre73},
where essentially linear flow profiles were found for bubble
rafts, i.e., in absence of a top plate.
\par

By comparing the results obtained from the velocity profiles with
the rheometrical measurements, we note a remarkable difference
between the scaling of the bubble-bubble drag forces at the bubble
level, which we have mimicked by strictly ordered bubble rheology,
and the scaling of the averaged forces at the bulk level, which we
have extracted from the velocity profiles and confirmed by
rheometry: we find $F_{bb} \sim (\Delta v)^{0.70}$ at the bubble
level and $\overline{F}_{bb} \sim (\Delta v)^{0.36}$ at the bulk
level.

We speculate that this is closely connected  to the non-affine
behavior of the bubbles \cite{durian,liuletter, ellenbroek}: close
to the jamming transition, the effective viscosity of the foam
becomes anomalously large due to the fact that bubble motion is
much more complicated than if the bubble motion would have been
affine, i.e., where the bubbles follow the imposed shear
\cite{liuletter}. This picture is corroborated by recent
simulations on the bubble model \cite{durian}, where one recovers
this ``renormalization'' of the drag force exponent
\cite{olsson,langlois,remmers}. The precise microscopic mechanism, though, is far from understood.
\par

%
%

\par
\begin{figure}[t]
\includegraphics[width=\columnwidth]{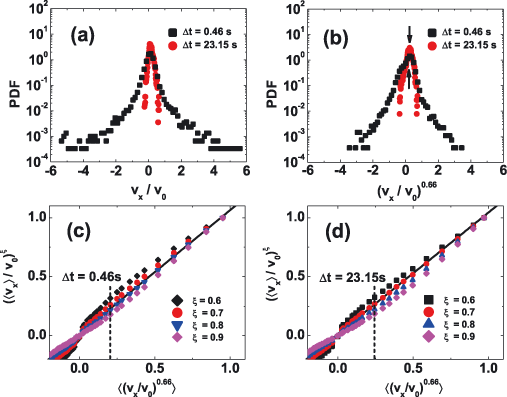}
\caption{(Color online) Dimensionless velocity distributions
measured for $W=7$ cm, $v_0=0.25$~mm/s, and for a $y$-position 19
mm away from the center of the gap. Here the averaged velocity
equals $3.1\times 10^{-2}$ mm/s and the local strain rate equals
$5.6\times 10^{-3}$ s$^{-1}$. (a) Distribution of $v_x/v_0$ for a
short time interval ($\Delta t =0.46$ s, black squares) and longer
time interval ($\Delta t =23.15$ s, red circles). For the averages
of the dimensionless velocity distributions we find $\langle
v_x/v_0 \rangle \approx 0.125$, independent of the time averaging
interval. (b) Distribution of $(v_x/v_0)^{2/3}$ for a short time
interval ($\Delta t = 0.46$ s, black) and longer time interval
($\Delta t = 23.15$ s, red). The averages of the scaled
dimensionless velocity distributions equal $\langle
(v_x/v_0)^{2/3} \rangle_{0.46 s} \approx 0.205$ and $\langle
(v_x/v_0)^{2/3} \rangle_{23.15 s} = 0.245$. The significance of
this is that $\langle v_x/v_0 \rangle^{2/3} \approx 0.248$, which
is significantly better approximated by the longer time average.
(c) Comparison of $\langle v_x/v_0 \rangle ^{\xi}$ and $\langle
v_x/v_0 \rangle^{2/3}$ along the flow profile for $\Delta t =
0.46$ s, and for four values of $\xi$ as indicated. The best
linear relation is obtained for $\xi \approx 0.80$. Dotted
vertical line indicates the averages shown in panel (b). (d) Same
as (c), now for $\Delta t = 23.15$ s. The best linear relation is
obtained for $\xi \approx 0.72$. } \label{fluct}
\end{figure}

One may wonder why the modification of the exponent of the drag
force law is strong for the inter-bubble forces but weak or
essentially absent for the bubble wall drag forces. We have no
definite answer, although we are fairly confident that the
bubble-wall drag forces indeed are not modified. We base this
assertion on explorations of the bubble trajectories, described
below.
\par
If we assume the Bretherton expression, Eq.~(\ref{fbb}), to be the
correct expression that gives the instantaneous bubble-wall drag
force as a function of the instantaneous bubble velocity, our
claim is that the averaged bubble-wall drag forces scale similar
to the individual bubble-wall drag force:
\begin{equation} \label{order_av}
\langle(\vec{v}/|v|)_x |v|^{2/3}\rangle \approx \langle v_x \rangle^{2/3}~.
\end{equation}
Hence we claim that the time averaged bubble wall drag force is
proportional to $\langle v_x \rangle^{2/3}$, which is the
expression we employ in our model to estimate $\overline{F}_{bw}$.
In other words, we can interchange the order of taking time
averages and ``raising to the power 2/3".

To check this,  we have performed accurate bubble tracking
 and calculated and compared \emph{$\langle
F_{bx}\rangle \equiv \langle(\vec{v}/|v|)_x |v|^{2/3}\rangle$} and
$\langle v_x \rangle^{2/3}$ \cite{mobius,notetrex}. In
Fig.~\ref{fluct}a-b we show examples of distributions of both
$\langle F_{bx}\rangle$ and $\langle v_x \rangle^{2/3}$, based on
short and long time velocity estimates at a fixed position in the
cell. For long times these distributions are narrower and have
less weight around zero. For the examples shown in
Fig.~\ref{fluct}a-b, the averages of the dimensionless velocity
distributions equal $\langle v_x/v_0 \rangle \approx 0.125$,
independent of the time averaging interval. Hence, $\langle
v_x/v_0 \rangle^{2/3} \approx 0.248$. The averages of the
distributions of $(v_x/v_0)^{2/3}$, taken over different time
intervals, depend now on this time interval and approximate
$\langle v_x/v_0 \rangle^{2/3} \approx 0.248$ better the longer
the time interval is: we find $\langle (v_x/v_0)^{2/3}
\rangle_{0.46 s} \approx 0.205$ while $\langle (v_x/v_0)^{2/3}
\rangle_{23.15 s} = 0.245$. Since the drag force model deals with
(long) time averages, the improvement of the agreement with time
is encouraging.

The connection between $\langle F_{bx}\rangle$ and $\langle v_x
\rangle^{2/3}$ can be probed in more detail by plotting $\langle
F_{bx}\rangle$ as function $\langle v_x \rangle^{\xi}$ for a range
of strain rates, and estimating for which value of $\xi$ these two
quantities are proportional. The data in Fig.~\ref{fluct}c shows
that for short times, a value of $\xi \approx 0.80$, significantly
different from 2/3, leads to the best correlation, while for
longer times --- \ref{fluct}(d) ---, the best value is $\xi \approx
0.72$. Therefore, the longer the time interval, the closer $\xi$
approaches 2/3. The underlying reason is that for increasing time
intervals, the distribution of $v_x/v_0$ becomes narrower and
narrower and peaked away from zero, and thus we indeed can
interchange the order of taking time averages and ``raising to the
power 2/3".

\par
The bubble-bubble drag forces, on the other hand, involve velocity
{\em differences}, and even at long times we expect their
probability distribution to have significant weight around $\Delta
v =0$. The situation is then qualitatively similar to that shown
in Fig.~\ref{fluct} for short times, and a change from local to
global exponent appears reasonable. Unfortunately, testing this
explicitly in our data for the bubble trajectories has proven to
be prohibitively difficult, not only because velocity differences
are smaller and more noisy than velocities, but also since bubble
contacts are very hard to establish unambiguously. The precise
mechanism responsible for the ``renormalization'' that leads to
the exponent $\beta \approx 0.4$ remains therefore open.
\par

Finally, the origin of the edge effects that prevent us from
fitting our full experimental curves with the model profiles,
might be due to the fluid drag near the wheels that was discussed
in section II C. Alternatively the origin might lie in the absence
of a local flow rule near the driving wheels as reported in
\cite{collin}. One way to resolve this is accommodating non-local
behavior in our model, for instance by incorporating drag terms
due to next nearest lanes, similar to the cooperativity length
introduced in \cite{collin}. We have not pursued this avenue.

\section{Ordered foams}\label{sec:ord}
We have postulated  that the disordered bubble motion underlies
the anomalous relation between the local bubble-bubble drag forces
and the global viscous stresses. To corroborate this conjecture,
we shear ordered, monodisperse foams in the linear geometry,
similar to what was done in \cite{denninpre73}. In this case the
bubbles are expected to move affinely  with the global shear, in
which case one would expect the global viscous drag forces to
scale the same as the local ones.
\par
\begin{figure}[tb]
\includegraphics[width=8cm]{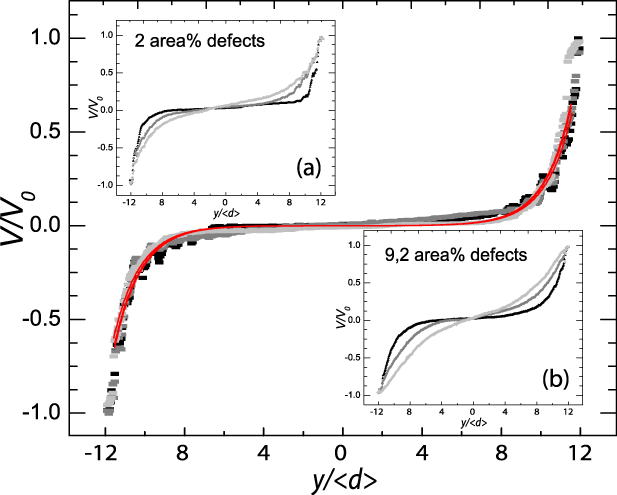}
\caption{(Color online) (a) Velocity profiles for a monodisperse,
ordered foam with the crystal axis aligned with the wheels. Gap $W$ =
7 cm and $v_0= 0.083$ (black), 0.26 (dark grey) and 0.83 (light
grey) mm/s. Solid curves indicate fits to the model
Eq.~(\ref{modeleq}) with $k =0.3$, $ \beta =2/3$. (a-b) Velocity
profiles for an ordered foam consisting of 2.7 mm bubbles for same
drving velocities as main panel, to which defects are added in the
form of an increasing area fraction of 1.8 mm bubbles as
indicated.} \label{rateinddep}
\end{figure}
We shear a monodisperse, ordered foam with bubbles of size 2.7 mm,
produced by blowing nitrogen through one syringe needle at fixed
flow rate, at a gap $W$ of 7 cm at $v_0= 0.083, 0.26$ and $0.83$
mm/s. We recover the rate independent and strongly shear banded
velocity profiles reported in \cite{denninpre73} (see
Fig.~\ref{rateinddep}). As in the case of the bidisperse foams, we
fit model profiles to our experimental data. For our model to
yield  rate independent velocity profiles,  the drag forces need
to balance in the same ratio for all driving velocities. This can
only be achieved if $\beta=2/3$ since we have already confirmed
with rheometry that the exponent governing bubble-wall drag is
$2/3$. Indeed we find that the experimental profiles are best fit
by model profiles if one fixes $k=0.3$ and $\beta=0.67\pm 0.05$
\cite{footnote2}, see Fig.~\ref{rateinddep}.

\subsection{Disorder}
In our experiment, the complex bubble motion is closely connected
to the anomalous scaling of the bubble-bubble drag force, which in
turn is reflected in the observed rate dependence of the velocity
profiles. We can thus investigate for which levels of disorder the
rate dependence of the velocity profiles occurs  by gradually
increasing the disorder, starting from a monodisperse foam.
\par
To this end we record velocity profiles in a monodisperse foam
made of 2.7 mm size bubbles in which we gradually increase the
area fraction of smaller (1.8 mm) bubbles. After mixing the two
species we  measure velocity profiles at $v_0 = 0.083, 0.26$ and
$0.83$ mm/s. We already observe the occurrence of rate dependent
velocity profiles for small quantities of defects, see inset (a)
and (b) of Fig.~\ref{rateinddep}, and by visual inspection, we
already see the swirling patterns, typical of our 41/59 bidisperse
foam, occurring at 2 \% disorder. These findings indicate that rate
independent flows are in fact limited to a narrow region close to
the almost singular case of completely ordered foams.

\section{Role of the packing fraction}\label{sec:pac}

In this section we will discuss linear shear experiments where we
will vary both the packing fraction (or \emph{wetness}) of our
foam $\phi$ as well as the applied strain rate, to investigate the
flow behavior of these foams as a function of density. In
particular, we will closely approach the jamming transition,
located at $\phi_c \approx 0.84$. This allows us to test our drag
force balance model over a wide range of experimental situations.
Our main findings are that, first, the scaling exponent $\beta$
appears to be independent of $\phi$, and second, that the
pre-factor $k$ is our model (Eq.~\ref{modeleq}) varies as
$1/(\phi-\phi_c)$ where $\phi_c \approx 0.84$.

\subsection{Varying and measuring $\phi$}
In order to vary $\phi$, we vary the vertical gap  between the
glass plates and the bulk solution between 3 and 0.2 mm. We do
this by adding or retracting fluid from the reservoir. For large
gaps the bubbles get stretched in the vertical direction, and
share large deformed facets --- the foam effectively becomes dry.
For small gaps the bubbles acquire a pancake-like shape, close to
purely disc like in the horizontal plane, with only small facets
between neighboring bubbles --- the foam effectively becomes wet.

\begin{figure}[tb]
\includegraphics[width=80mm]{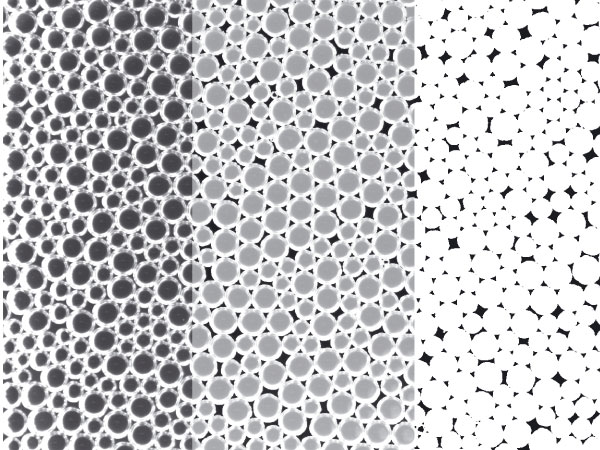}
\caption{Image manipulations leading to a definition of $\phi$.
Left: Raw image. Center: Raw image with reconstructed bubble areas
superposed. Note the good agreement. Right: Final binarized image
from which packing fraction is deduced.} \label{wetness}
\end{figure}

To create a homogeneous gap between the liquid surface and the
glass plate, we place additional supports under the glass plate to
prevent sagging of the top plate during the runs. We monitor the
gap width with a Mitutoyo digital depth gauge. If the gap becomes
smaller than 0.2 mm the bubbles unjam \cite{coxpm2008, footnote3}.

We find that in the linear shear cell the accessible range in
$\phi$ is $0.86 \lesssim \phi \lesssim 0.97$. If we stay between
these limits the system we study is jammed and quasi
two-dim\-en\-si\-o\-nal. It should be noted that for the runs
performed at fixed wetness, discussed in the previous sections, we
find $\phi = 0.965 \pm 0.005$, in reasonable agreement with
previous reports on the maximum $\phi$ that can be obtained in our
type of setup \cite{raufaste}.

Determining the liquid fraction is not trivial, since various
horizontal cuts through the bubble layer will yield different
values \cite{wetnote}. We choose our lighting of the bubbles such
that the contacts between adjacent bubbles are optimally resolved.
We then extract $\phi$ through image analysis, as illustrated in
Fig.~\ref{wetness}. We first binarize the images, after which both
the bubble centers and the interstices appear bright. We remove
the interstices by morphological operations. We then invert the
binarized image and fill up the remaining bubble contours. We have
checked that the resulting bright disc optimally matches the
original bubble contour, see Fig.~\ref{wetness}. We then calculate
the ratio of white pixels over the total number of pixels and
hence obtain a reasonable estimate of $\phi$.

Now that we have obtained good estimates of the packing fraction
$\phi$, we can probe the role of the wetness in setting the flow.
We first, in section \ref{secalpha}, briefly discuss a local probe
of the non-affine motion, which shows that the bubble motion
becomes increasingly non-affine when the wetness is increased. We
then investigate the variation of the flow behavior with $\phi$,
using our model Eq.~(\ref{modeleq}). We first establish, in
section \ref{secsubbeta}, that the exponent $\beta$ does not vary
with $\phi$ --- surprising, give the varying degree of
non-affinity. We then find, in section \ref{secsubk}, that the
force pre-factor $k$ varies strongly with $\phi$ and vanishes at
$\phi_c \approx 0.84$ as $1/(\phi-\phi_c)$.

\begin{figure}[tb]
\begin{center}
\includegraphics[width=7cm]{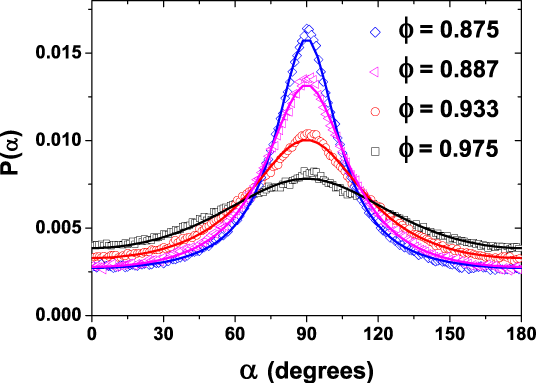}
\caption{(Color online)
Displacement angle distributions
$P(\alpha)$ for runs for which $v_0=0.26$ mm/s, $W=5$ cm and $\Delta t=0.46$~s
averaged over the shear banded region (
$0<y< W/3$ and $2W/3<y<W$) for the range of packing fractions as indicated.
 } \label{figalpha}
\end{center}
\end{figure}

\subsection{A local measure of the non-affine bubble
motion: $P(\alpha)$}\label{secalpha}

A crucial feature of deformations found in simulations of packings
of frictionless discs near jamming is the strongly non-affine
nature of the particle (bubble) motion
\cite{durian,liuletter,makseprl1999, ellenbroek}.
Recently, a simple local probe of this affinity was introduced by
Ellenbroek {\em et al.} who performed simulations of soft
frictionless discs \cite{ellenbroek}. Defining the displacements of
contacting particles $i$ and $j$ as $\vec{u}_i$ and $\vec{u}_j$,
and the vector that connects the centers of particles $i$ and $j$
as $\vec{r}_{ij}$, the relative displacement angle $\alpha$ was
defined as the angle between $\vec{r}_{ij}$ and $\vec{u}_i -
\vec{u}_j$. In other words, $\alpha =0^{\circ}$ corresponds to
particles moving away from each other, $\alpha =180^{\circ}$
corresponds to particles moving closer, and $\alpha =90^{\circ}$
corresponds to particles sliding past one another.

The probability distribution $P(\alpha)$ was found, for shear
deformations in particular, to be well fitted by a (periodically
extended) Lorentzian peaked around $90^{\circ}$
\cite{ellenbroek,wouterlongpreinpreparation}. The width of the peak
scales with distance to jamming --- at jamming, $P(\alpha)$
approaches a delta function peaked at $\alpha=90^{\circ}$.

Of course, in our experiment we have flow, and we cannot determine
deformations in linear response. Moreover, our system is not
homogeneous. Nevertheless, as a coarse measure of the degree of
non-affine bubble motion, which we claim underlies the anomalous
scaling exponent $\beta$ in disordered systems, we have calculated
$P(\alpha)$ focussing on finite time displacement fields
($v_0=0.216$ mm/s, $W=5$ cm, $\Delta t=0.46$~s).

In Fig.~\ref{figalpha} we show
$P(\alpha)$ averaged over the regions $0<y<W/3$ and $2W/3<y<W$ where most of the
flow takes place, and averaged in the $x$-direction  over 50 mm in the center
of the cell. We limit ourselves to this region, because,
in particular for the wet runs, there is hardly any
flow in the center region of the cell and the peaks in $P(\alpha)$ are less
pronounced in this region.
We find that, analogous to what is found in
simulations \cite{ellenbroek}, the distributions become
increasingly peaked around $\alpha=90^{\circ}$ for increasing
wetness. Moreover, the distributions are well fit by the same
Lorentzian fit that also captures the numerical displacement
fields well \cite{wouterlongpreinpreparation}.

Hence, this simple measure of non-affine motion strongly indicates
that the degree of non-affinity increases for wetter foams. We
believe that this is the first experimental measurement of this
distribution that shows the proximity of the jamming transition.
Detailed studies of the role of the local strain rate or the time
interval over which displacements are measured are deferred to
later work.

\begin{figure}[tb]
\begin{center}
\includegraphics[width=6
cm]{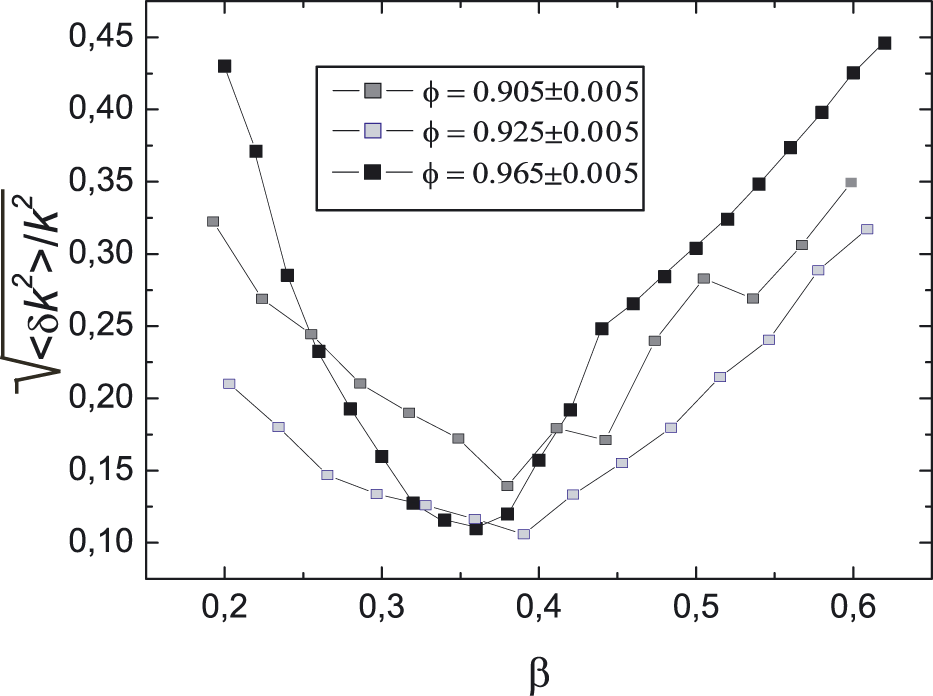} \caption{(a) variance in $k$ values for all
six runs performed at $\phi =0.905$ (grey squares) and $\phi
=0.925$ (light grey squares). The variance at $\phi =0.965$ (black
squares) is data from Fig.~\ref{ratedep}(f). A clear minimum can
be observed around $\beta=0.38$.} \label{dkvsphitot}
\end{center}
\end{figure}

\subsection{Variation of the exponent $\beta$ with $\phi$}\label{secsubbeta}

We now investigate the validity of applying the drag force balance
model with a fixed $\beta=0.36$ for varying $\phi$. The
microscopic exponent 2/3 which governs the flow of a bubble past a
wall appears to be independent of the particularities of the foam
flow \cite{cantat, raufastephd}. On the other hand, it is not at
all obvious that $\beta$, which governs the averaged bubble-bubble
drag forces, does not depend on $\phi$. As we have seen, $\beta$
is set by the disorder in the system and the non-affine bubble
motion that occurs in conjunction with that, and as we have shown
in the previous section, the degree of non-affinity varies
substantially with $\phi$.

To see if $\beta$ indeed depends on the foam density we perform
two additional scans over the same six shear rates as employed in
section III for a bidisperse foam  at a gap width $W=7$~cm, while
first fixing $\phi =0.905 \pm 0.005$ and then $\phi =0.925 \pm
0.005$. We look for a minimum of the variance in $k$ over the six
velocity profiles as a function of $\beta$ (see grey and light
grey squares in Fig.~\ref{dkvsphitot}). We observe that the model
fits best to all six runs performed at $\phi =0.905$ for $\alpha =
2/3, \beta = 0.38 \pm 0.05$ (see Fig.~\ref{dkvsphitot}) and $k
=7.5$, whereas the model best matches the runs performed at $\phi
=0.925$ for $\alpha = 2/3, \beta = 0.39 \pm 0.05$ (see
Fig.~\ref{dkvsphitot}) and $k =5.8$, thus strongly indicating that
within our range of accessible liquid fractions $\beta$ seems to
be constant. For comparison, we include the variance for the runs
described in section III B  that was plotted in
Fig.~\ref{ratedep}(f). Remarkably, $\beta$ remains a constant with varying $\phi$
while the degree of non-affinity varies.
While we do not pretend to understand this, we do remark that $\beta$ and $P(\alpha)$
essentially encode different routes towards jamming and thus towards increasing non-affinity: $\beta$ is renormalized by the increasing non-affinity as one lowers the strain rate $\dot{\gamma}$ towards jamming, while $P(\alpha)$ monitors non-affinity as a function of density.

\subsection{Scaling of the force pre- factor $k$ with $\phi$}
\label{secsubk}

\begin{figure}[tb]
\begin{center}
\includegraphics[width=6
cm]{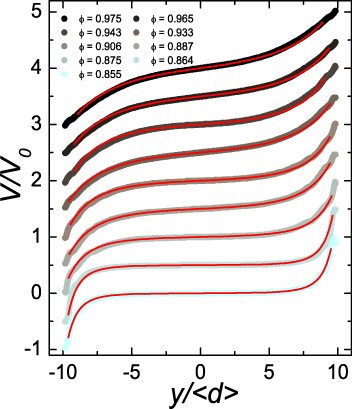} \caption{(Color online) Velocity
profiles in linearly sheared foam at fixed driving rate ($v_0 =
0.26$~mm/s), for $\phi$ varying between $0.855 \leq \phi \leq
0.975$ at $W=7$~cm. Fits are solutions to linear drag force
balance model with $\alpha=0.67$ and $\beta = 0.36$ fixed. $k$ is
extracted from the fits and plotted in Fig.~\ref{dkvsphitotII} as
a function of $\phi-\phi_c$. } \label{phirun}
\end{center}
\end{figure}

Now that we have established that $\beta$ is independent of
$\phi$, we will probe the variation of $k$ with $\phi$. We measure
averaged velocity profiles at gap widths $W = 5$~cm and $W = 7$~cm
and fixed $v_0 = 0.26$ mm/s (the third slowest driving
velocity), for packing fractions varying between $\phi=0.855$ and
$\phi=0.975$.  The velocity profiles  for $W=5$~cm are plotted in
Fig.~\ref{phirun}, and are seen to become increasingly shear
banded as we approach $\phi_c$ \cite{foot2}. This trend is reflected in the
increase of $k$ as we approach $\phi_c$. We obtain $k$ by  fitting
solutions of our drag force balance model with $\alpha=0.67$ and
$\beta = 0.36$ to these profiles. The resulting fits are shown as
red lines in Fig.~\ref{phirun}, and fit the data well.

\begin{figure}[tb]
\begin{center}
\includegraphics[width=6
cm]{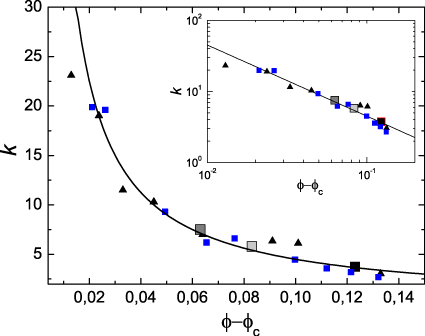} \caption{(Color online)  Scaling of $k$ with
$\Delta \phi \equiv \phi-\phi_c$. Triangles: data obtained from
fits depicted in Fig.~\ref{phirun} where $W$ = 5 cm. Squares: data
for gap of 7 cm.  Large squares correspond to runs at $v_0$ =0.26
mm/s from Fig.~\ref{dkvsphitot}. Solid line: $0.45/\Delta \phi$.
Inset: same data on log-log scale. } \label{dkvsphitotII}
\end{center}
\end{figure}

In Fig.~\ref{dkvsphitotII} we plot $k$ as a function of
$\phi-\phi_c$, with $\phi_c$ the theoretically predicted and
experimentally measured value of the unjamming packing fraction:
$\phi_c = 0.842$ \cite{olsson,bolton, lechenault}.  In good
approximation we obtain that
\begin{equation}
k \propto 1/(\phi-\phi_c)
\end {equation}

\par

We can tentatively explain the observed scaling of $k$ with a
simple argument based on the sizes of the facets in the foam. At
fixed $\phi$, our drag force balance model yields a value of $k$
that sets the relative influence of the bubble-wall drag with
respect to the bubble-bubble drag and which we have conjectured to
be given by $k \propto f_{bw}/f_{bb}$. As we have already
discussed, $f_{bw} \propto \sigma r_c$ with $r_c$ the radius of
the flattened contact between the bubble and the wall and $f_{bb}
\propto \sigma \kappa^2_c$, with $\kappa_c$ the radius of the
flattened contact between neighboring bubbles. Thus we expect:
\begin{equation}
k \propto r_c/\kappa_c^2. \label{kscaling}
\end{equation}
While $r_c$ is set by the buoyancy and hence does not vary
strongly with the gap distance between glass plate and liquid
surface --- only becoming slightly smaller as the bubbles get
stretched at large gaps --- $\kappa_c$ is strongly dependent on
the gap size and hence on the packing fraction of the foam.

The size of $\kappa_c$ should depend on the deformation (also
called the overlap) $\delta\xi$ as \cite{lacasse}:
\begin{equation}
\kappa_c \propto (\delta\xi)^{1/2}.
\end{equation}
Similar to simulations of two-dim\-en\-si\-o\-nal frictionless
discs \cite{ohern,ellenbroek} we can relate the overlap $\delta
\xi$ to the packing fraction $\phi$:
\begin{equation}
\delta\xi \propto \Delta \phi.
\end{equation}
\par
Simple substitution of this result into Eq.~(\ref{kscaling})
yields
\begin{equation}
k \propto r_c/\kappa_c^2 \propto 1/\delta\xi  = 1/(\Delta \phi),
\end{equation}
which is fully consistent with our experimental results, see the
solid line in Fig.~\ref{dkvsphitot}
\par
Note that in the above we have only focussed on the radius of the
deformed facets. A proper analysis would include the size of the
Plateau border around the contact, which is where the dissipation
also occurs \cite{denkov2,terriac}. For instance, in
\cite{raufastephd} the bubble-wall drag force scales as $F^{bw}
\propto Ca^{0.64}\phi_l^{-0.26}$ and a proper treatment would
entail such analysis, even \linebreak though the functional
dependence on the Plateau border size is always weak. Moreover, in
all of these works, the functional dependence of the drag force
with $\phi$ is smooth around $\phi_c$ and hence will not influence
the observed scaling around that point.

\section{Discussion and conclusion}\label{sec:con}
We have measured velocity profiles in linearly sheared
quasi-two-dimensional foams in the liquid-glass configuration. We
find that bidisperse, disordered foams exhibit strongly rate
dependent and inhomogeneous (shear banded) velocity profiles,
while monodisperse, ordered foams are also shear banded, but
essentially rate independent. We capture these findings in a simple
model that balances the drag forces in our system. The scaling
forms for these drag forces are verified by independent
rheological measurements. Finally, we apply our model to velocity
profiles obtained for foams at varying packing fraction, and
measure and describe the scaling of the inverse foam consistency
with packing fraction.

This work raises several questions. First, can the difference
between the local bubble-bubble drag force scaling and the global
(averaged) bubble-bubble drag force scaling be understood
theoretically? This difference in scaling exponents appears
similar to the change from local drag forces to global rheological
laws, observed in simulations of (variants) of the bubble model
\cite{durian,olsson,hatano,remmers,coreyPRE}, but a precise
connection is lacking at present. Closely connected, is our
scenario an example of a general route by which aspects of the
ubiquitous Herschel-Bulkley (power law) rheology observed for a
wide range of disordered materials can be rationalized?

Second, how robust are our experimental results? For example,
would similar flows in Hele-Shaw cells behave differently, as
suggested by the results of Debr\'{e}geas \cite{debregeasprl87}?
We also wonder if our model is able to capture shear banded flows
in Couette geometries, where the curvature plays an important
role, in particular since the foam has a finite flow threshold
\cite{denninreview}. Third, can similar phenomena and models as
described here be extended to three dimensional flows of foams and
emulsions --- where flows in the latter can be captured by
confocal imaging and MRI \cite{brujic,collin, ovarlez, rodts}?
Fourth, how should our local models be compared to the non-local
effects recently discussed for emulsion flows \cite{collin,
ovarlez}?
\vspace{-0.8cm}
\begin{acknowledgments}
\vspace{-0.2cm}
The authors wish to thank Jeroen Mesman for technical assistance.
GK kindly acknowledges Nikolai Denkov for illuminating
discussions. GK and MM acknowledge support from physics foundation
FOM, and MvH acknowledges support from NWO/VIDI.
\end{acknowledgments}


\begin{thebibliography}{99}
\bibitem{foot1} This is done to ensure the packing
fraction remains constant during the strain rate sweep; when the entire circumference of the wheel is
covered with bubbles a balance results between bubbles dragged out of the system and injected back in.
If we would sweep from slow to fast driving rates, this balance is not achieved, resulting in a
packing fraction that decreases during the experiment.
\bibitem{notetrex}
One subtlety one encounters is that from the tracking we can only
estimate the velocity from ratios of displacements and time
intervals, and for the experimentally accessible timescales, the
statistics of the velocities thus obtained have been found to
depend on the time interval
--- see \cite{mobius}.
\bibitem{footnote2} The value of $k$ is remarkably small. If we assume that prefactor $f_{bw}$ for the bubble wall
drag remains unchanged for the ordered foam, this means
that the bubble-bubble drag prefactor $f_{bb}$  is much larger compared to its value for a disordered foam.
Note however, that the power law exponent $\beta$ greatly influences the value of the drag force: for instance, if $\Delta v =0.001$~m/s,
then $(\eta v/\sigma)^{2/3} = 1.6 \times 10^{-3}$, whereas $(\eta v/\sigma)^{0.36} = 3.1 \times 10^{-2}$, which is more than an order of magnitude larger.
\bibitem{footnote3} This might be due to the fact that the gap is then of the size of the Plateau borders that connect the flat film between the bubble and the glass plate and the flat film between neighboring bubbles, and hence the latter vanishes. If the gap becomes larger than 3 mm the foam buckles and develops a three dimensional structure.
\bibitem{wetnote}
One could try to relate the liquid fraction to the gap between the
liquid surface and the glass plate. This distance, however, does
not unambiguously set $\phi$ in our experiment: we observe a large
hysteresis effect, i.e., increasing or decreasing the gap to a
certain value does not yield the same packing fraction $\phi$.
This probably due to the open boundary conditions in the
horizontal plane. Another measure that has been derived in
\cite{raufaste} relates the measured length of the deformed facets
of the bubbles just before a T1 event to $\phi$. In our
experiments, though, it is not clear how the occurrence of
T1-events can precisely be defined, since there is no obvious
separation of the deformation scales during and outside of a
T1-event.
\end{thebibliography}
\end{document}